\documentclass[aps,prd,superscriptaddress,twoside,twocolumn,nofootinbib,preprintnumbers]{revtex4}

\usepackage{amsmath}
\usepackage{graphicx}

\newcommand{\lesssim}{\mathrel{\mathpalette\vereq<}}

\begin{document}

\title{Walking from 750 GeV to 950 GeV in the Technipion Zoo}

\author{Shinya Matsuzaki}\thanks{{\tt synya@hken.phys.nagoya-u.ac.jp}}
      \affiliation{ Institute for Advanced Research, Nagoya University, Nagoya 464-8602, Japan.}
      \affiliation{ Department of Physics, Nagoya University, Nagoya 464-8602, Japan.}    
\author{Koichi Yamawaki} \thanks{{\tt yamawaki@kmi.nagoya-u.ac.jp}}
      \affiliation{ Kobayashi-Maskawa Institute for the Origin of Particles and the Universe (KMI) \\ 
 Nagoya University, Nagoya 464-8602, Japan.}

\date{\today}

\begin{abstract} 
If the 750 GeV diphoton excess is identified with the color-singlet isosinglet-technipion, $P^0$ (750),
in the one-family walking technicolor, as in our previous paper, then there should exist another color-singlet
technipion, isotriplet one,  $P^{\pm,3}$, definitely predicted at around 950 GeV independently of the dynamical details. 
The $P^{\pm,3}(950)$ are produced at the LHC via vector boson and photon fusion processes,
predominantly decaying to $W \gamma$, and $\gamma\gamma$, respectively. 
Those walking technicolor signals can be explored at the Run 2, or 3, 
which would further open a way to a plethora of yet other (colored) technipions.

\end{abstract}
\maketitle


The ATLAS and CMS groups~\cite{ATLAS-CONF-2015-081,CMS:2015dxe} have reported 
a diphoton excess with the global significance of about 3 standard deviations     
at around 750 GeV. 
It would provide a clue for new physics beyond the standard model.

In the previous work~\cite{Matsuzaki:2015che} the authors gave an interpretation for 
the diphoton excess by identifying the 750 GeV resonance as a color-singlet isosinglet technipion, 
$P^0$ 
of the one-family model~\cite{Farhi:1980xs},  which was shown~\cite{Jia:2012kd,Kurachi:2014xla} 
to have mass of this large in the walking technicolor having the large anomalous dimension
$\gamma_m=1$~\cite{Yamawaki:1985zg}. 
In this paper, we present another implication following the 750 GeV resonance: 
that is the presence of the technipion with the mass of 950 GeV, which is color-singlet isotriplet 
(denoted as $P^{\pm,3}$), enrolled in the technipion ``zoo" with sixty entries in total.

The one-family walking technicolor is a scale-invariant (walking) version of the original 
one-family technicolor model~\cite{Farhi:1980xs} a naive-scale up of QCD. 
The theory possesses eight technifermion flavors, $F=(Q^c, L)$, 
which consists of six  techniquarks $(Q^c=(U,D)^c)$ having the QCD charge $(c=r,g,b)$ and two technileptons $(L=(N,E))$, singlet under the QCD. 
The chiral symmetry in the theory is thus enlarged from the $SU(2)_L \times SU(2)_R$ in the standard model 
to the $SU(8)_L \times SU(8)_R$.  
The technifermions develop the chiral condensate $\langle \bar{F}F \rangle$ by the strong dynamics 
to break the chiral $SU(8)_L \times SU(8)_R$ symmetry down to the vectorial $SU(8)_V$. 
The sixty-three Nambu-Goldstone (NG) bosons then emerge, among which three are eaten by the $W$ and $Z$ bosons 
once the electroweak gauge is turned on, while other sixty become pseudo-NG bosons due to the explicit breaking 
effects supplied outside the walking technicolor dynamics.   
Thus the low-lying spectra consist of 
those sixty technipions, 
as well as the characteristic composite Higgs ( ``technidilaton'', a pseudo NG boson of the scale symmetry, 
predicted in the walking technicolor~\cite{Yamawaki:1985zg,Bando:1986bg}), identified as the 125 GeV LHC Higgs. 
(Several discussions on the lightness of the technidilaton and 
the consistency of its coupling property with the LHC Higgs have been given in recent works.  
See Refs.~\cite{Matsuzaki:2012gd,Aoki:2014oha}.)

The technipions are classified on the basis of the standard model charges: 
the color-singlet technipions $P^0$ and $P^i$ (with $i=1,2,3$ being the isospin charges) 
are constructed from technifermions as 
$P^0 \sim \frac{i}{4\sqrt{3}} (\bar{Q} \gamma_5 Q - 3 \bar{L} \gamma_5 L)$, 
$P^i \sim \frac{i}{4\sqrt{3}} (\bar{Q} \gamma_5 \sigma^i Q - 3 \bar{L} \gamma_5 \sigma^i L)$, 
where $\sigma^i$ stands for the Pauli matrices.  
As was discussed in Refs.~\cite{Jia:2012kd,Kurachi:2014xla} in the context of the walking technicolor, 
they get the masses due to a four-fermion interaction induced by an extended technicolor which explicitly breaks the 
associated chiral symmetry (but keeps the standard-model symmetry), 
\begin{equation} 
 \frac{1}{\Lambda^2_{\rm ETC}} 
 \left( 
 \bar{Q}Q \bar{L}L - \bar{Q} \gamma_5 \sigma^i Q \bar{L} \gamma_5 \sigma^i L 
\right) 
\,. 
\end{equation}
The masses are calculated by using the standard current algebra. Then one gets 
the formula~\cite{Jia:2012kd,Kurachi:2014xla},  
\begin{equation} 
 m_{P^i}^2 = \frac{8}{5} m_{P^0}^2 
 \,.  \label{mass:formula}
\end{equation}
Remarkable to note is that this formula is fixed without any detail of the walking dynamics 
and modeling of the extended technicolor: the prefactor $(8/5)$ has merely come from the difference 
in the associated chiral charges for $P^{0}$ and $P^{\pm,3}$.  
 As was shown in Ref.~\cite{Matsuzaki:2015che}, the $P^0$ can be interpreted as 
 the 750 GeV diphoton resonance, so we take $m_{P^0}=750$ GeV in Eq.(\ref{mass:formula}) 
 to get the $P^i$ mass:  
\begin{equation} 
 m_{P^i} = \sqrt{\frac{8}{5}} m_P^0 \Bigg|_{P^0 \equiv P^0(750)} \simeq 950 \,{\rm GeV} 
\,.  \label{mass:950}
\end{equation} 
Thus the presence of the 750 GeV resonance simultaneously predicts the 950 GeV isotriplet technipion, $P^i \equiv P^i(950)$.

Besides the color-singlet technipions, the theory predicts the color-octet and -triplet ones. 
The masses of the colored technipions are originated from a different source: 
those are generated by the QCD interactions, just like the photon exchange contribution to the charged pion mass in QCD. 
The explicit breaking effect of all the technipions is actually amplified by 
the large anomalous dimension $\gamma_m\simeq 1$ characteristic to the walking technicolor~\cite{Jia:2012kd,Kurachi:2014xla} 
to lift the mass up to ${\cal O}$(TeV). The precise size of the mass is, however, 
subject to the nonperturbative calculation of the vector current correlator in 
the walking dynamics, in sharp contrast to the case of the color-singlet technipions, 
particularly the ratio  $m_{P^i}/ m_{P^0}$ which is free from the dynamical details as mentioned above.

The couplings of $P^i(950)$ to the standard-model gauge bosons are
only given by the Wess-Zumino-Witten term~\cite{Wess:1971yu}  for the non-Abelian anomaly of 
the underlying walking technicolor, 
since the three-NG-boson vertex is forbidden by the low-energy theorem of the spontaneously broken chiral symmetry in 
the non-anomalous part (See, e.g., Sec. 2.2. of Ref.~\cite{Bando:1987br}), 
which is in sharp contrast to the coupling of the (charged) non-NG boson-heavy Higgs boson in extended Higgs models. 
The Wess-Zumino-Witten construction for the chiral $SU(8)_L \times SU(8)_R$ symmetry reads 
\begin{equation} 
 S_{\rm WZW} = - \frac{N_C}{12 \pi^2 F_\pi} 
 \int_{M^4} 
{\rm tr}[ \left( 3 d {\cal V} d {\cal V} + d {\cal A} d {\cal A} \right) \pi  ] 
\,,  \label{WZW}
\end{equation}
which breaks the intrinsic parity~\footnote{
The intrinsic parity is defined to be even when 
a particle has the parity $(-1)^{\rm spin}$, otherwise odd}~\cite{Jia:2012kd},  where $N_C$ denotes the number of the technicolor 
and $F_\pi$ is the technipion decay constant, fixed by the electroweak scale $v_{\rm EW}=246$ GeV 
as 
\begin{equation} 
 F_\pi = v_{\rm EW}/\sqrt{N_D}\Bigg|_{N_D=N_F/2=4} = 123 \,{\rm GeV}
 \,, \label{Fpi}
\end{equation}
for the one-family model with the eight techni-flavors, forming the four electroweak doublets ($N_D=N_F/8=4$).  
Equation (\ref{WZW})  
has been written in terms of differential form. 
The $P^i(950)$ are parametrized in the $\pi$ matrix, $\pi \ni P^i X^i_P$, 
with the corresponding $SU(8)$ generator, 
\begin{equation} 
X_P^i 
= \frac{1}{4 \sqrt{3}} 
\left( 
\begin{array}{c|c} 
 \sigma^i \otimes {\bf 1}_{3 \times 3} & 0 \\ 
\hline 
 0 & - 3 \cdot \sigma^i 
\end{array}
\right)_{\bf 8 \times 8} 
\,. 
\end{equation} 
The standard-model gauge boson fields $(W_\mu^\pm, Z_\mu, A_\mu)$  
are embedded in the chiral-external gauge fields ${\cal V}_\mu$ and 
${\cal A}_\mu$ as follows~\cite{Jia:2012kd}: 
\begin{eqnarray} 
 {\cal V}_\mu 
 &=& 
 e Q_{\rm em} A_\mu + \frac{e}{2sc} \left( I_3 - 2 s^2 Q_{\rm em} \right) Z_\mu 
\nonumber \\ 
&& 
+ \frac{e}{2 \sqrt{2} s} \left( W_\mu^+ I^+ + W_\mu^- I^- \right) 
\,, \nonumber \\ 
{\cal A}_\mu 
&=& 
- \frac{e}{2 sc} I_3 Z_\mu 
-    \frac{e}{2 \sqrt{2} s} \left( W_\mu^+ I^+ + W_\mu^- I^- \right) 
\,,
\end{eqnarray} 
where $e$ is the electromagnetic coupling, 
$s$ $(c^2\equiv 1- s^2)$ denotes the weak mixing angle, and 
\begin{eqnarray} 
Q_{\rm em} 
&=& I_3 + Y 
\,, \nonumber \\
I_3 &=& 
\frac{1}{2}
\left( 
\begin{array}{c|c} 
\sigma^3 \otimes {\bf 1}_{3 \times 3}&  0 \\ 
\hline 
0& \sigma^3 
\end{array}
\right)
\,, 
\nonumber \\  
Y &=& 
\frac{1}{6}
\left( 
\begin{array}{c|c} 
{\bf 1}_{2\times 2} \otimes {\bf 1}_{3 \times 3}&  0 \\ 
\hline 
0& -3 \cdot {\bf 1}_{2 \times 2}  
\end{array}
\right)
\,, \nonumber \\ 
I^\pm 
&=& 
\frac{1}{2}
\left( 
\begin{array}{c|c} 
\sigma^\pm \otimes {\bf 1}_{3 \times 3}&  0 \\ 
\hline 
0& \sigma^\pm 
\end{array}
\right)
\,, 
\end{eqnarray}
with $\sigma^\pm = (\sigma^1 \mp i \sigma^2)$. 
In evaluating Eq.(\ref{WZW}) we have omitted the gluon field to which 
the $P^i(950)$ 
does not couple because of the isospin symmetry. 
From these, we extract the $P^i(950)$ couplings to find 
\begin{eqnarray} 
 {\cal L}_{P^3 AA} 
 &=& - \frac{e^2 N_C}{4 \sqrt{3} \pi^2 F_\pi} 
P^3 dAdA 
\,, \nonumber \\ 
 {\cal L}_{P^3 ZZ} 
 &=&  \frac{e^2 (c^2 - s^2) N_C}{8 \sqrt{3} \pi^2 c^2 F_\pi} 
 P^3 dZdZ  
\,, \nonumber \\ 
 {\cal L}_{P^3 AZ} 
 &=& - \frac{e^2 (1- 4 s^2) N_C}{8 \sqrt{3} \pi^2 sc F_\pi} 
P^3 dA dZ 
\,, \nonumber \\ 
 {\cal L}_{P^\pm AW} 
 &=& - \frac{e^2  N_C}{8 \sqrt{3} \pi^2\, s\, F_\pi} 
P^+ d A d W^- 
 + {\rm h.c.} 
\,, \nonumber \\ 
 {\cal L}_{P^\pm ZW} 
 &=& \frac{e^2  N_C}{8 \sqrt{3} \pi^2 \, c \, F_\pi} 
P^+ d Z dW^- 
 + {\rm h.c.}
\,, \label{P950-couplings}
\end{eqnarray}
where $P^\pm \equiv (P^1 \mp i P^2)/\sqrt{2}$ and $dV_1dV_2 \equiv \epsilon^{\mu\nu\rho\sigma} \partial_\mu V_{1\nu} \partial_\rho V_{2 \sigma}$ for 
arbitrary vector fields $V_{1\mu}$ and $V_{2\mu}$. 
Note the absence of the $P^3$ coupling to $WW$ due to the one-family $SU(8)$ symmetry, 
in a way that ${\rm tr}[X_P^3 \{I^+, I^-\}]=0$, 
where the contribution from techniquarks 
are canceled by that from techniletons, as in the case of 
the $P^0(750)$~\cite{Matsuzaki:2015che,Jia:2012kd}. 
   Thus, no coupling of $P^3(950)$ to $WW$ 
as well as $P^0(750)$ is the characteristic feature: if the 750 GeV resonance is 
in the future confirmed not only in the diphoton channel, 
but also in the $WW$ channel, the present one-family model will definitely 
be ruled out.

  From Eq.(\ref{P950-couplings}) we thus compute the partial decay rates of the $P^i(950)$ 
  to get 
\begin{eqnarray} 
&&  \Gamma(P^3 \to \gamma\gamma) 
\nonumber \\ 
&&  = 
  \left( \frac{\alpha_{\rm em} N_C}{\sqrt{3} \pi F_\pi}  \right)^2 \frac{m_{P(950)}^3}{16 \pi} 
 \,, \nonumber \\ 
&& \Gamma(P^3 \to ZZ) 
\nonumber \\ 
&& 
= 
  \left( \frac{\alpha_{\rm em} (c^2-s^2) N_C}{2\sqrt{3} \pi \, c^2 \,  F_\pi}  \right)^2 \frac{m_{P(950)}^3}{16 \pi} 
\left( 1 - \frac{4 m_Z^2}{m_{P(950)}^2} \right)^{3/2}  
  \,, \nonumber \\ 
&&    \Gamma(P^3 \to Z\gamma) 
\nonumber \\ 
&& 
= 
  \left( \frac{\alpha_{\rm em} (1-4 s^2) N_C}{2\sqrt{3} \pi \, sc \, F_\pi}  \right)^2 \frac{m_{P(950)}^3}{32 \pi} 
\left( 1 - \frac{m_Z^2}{m_{P(950)}^2} \right)^{3}  
\,, 
\end{eqnarray}  
  and 
\begin{eqnarray} 
&& \Gamma(P^\pm \to W^\pm \gamma) 
\nonumber \\ 
&& 
= \left( 
 \frac{\alpha_{\rm em} N_C}{2\sqrt{3} \pi \, s \, F_\pi}
\right)^2
\frac{m_{P(950)}^3}{32 \pi} 
\left( 1 - \frac{m_W^2}{m_{P(950)}^2} \right)^{3} 
\,, \nonumber \\ 
&& \Gamma(P^\pm \to W^\pm Z) 
\nonumber \\  
&& 
= 
\left( 
 \frac{\alpha_{\rm em} N_C}{2\sqrt{3} \pi \, c \, F_\pi}
\right)^2
\frac{m_{P(950)}^3}{32 \pi} 
\left( 1 - \left( \frac{m_W + m_Z}{m_{P(950)}} \right)^2 \right)^{3/2}
\nonumber \\  
&& 
\hspace{100pt} 
\times 
 \left( 1 - \left( \frac{m_W - m_Z}{m_{P(950)}} \right)^2 \right)^{3/2} 
\,, 
\end{eqnarray}
where $\alpha_{\rm em} \equiv e^2/(4\pi)$. 
Note that the branching ratios are estimated independently of $N_C$ and $F_\pi$ to be 
\begin{eqnarray} 
 {\rm Br}[P^3 \to \gamma\gamma] 
&\simeq& 89.5 \, \% 
\,, \nonumber \\  
  {\rm Br}[P^3 \to ZZ] 
&\simeq& 10.2 \, \% 
\,, \nonumber \\ 
  {\rm Br}[P^3 \to Z \gamma] 
&\simeq& 0.30 \, \% 
  \,, \label{BR:P3}
\end{eqnarray}
and 
\begin{eqnarray} 
  {\rm Br}[P^\pm \to W^\pm \gamma] 
&\simeq& 
77 \, \%  
\,, \nonumber \\ 
  {\rm Br}[P^\pm \to W^\pm Z] 
&\simeq& 23 \, \% 
\,. \label{BR:Ppm}
\end{eqnarray}

The total widths are estimated by using the value of $F_\pi$ in Eq.(\ref{Fpi}) 
and taking typical numbers for $N_C$, say, $N_C =3,4$: 
\begin{eqnarray} 
\begin{array}{c||c|c} 
& N_C=3 & N_C=4 \\ 
\hline 
\Gamma_{\rm tot}^{P^3(950)} [{\rm MeV}]  
& 23 &  42  \\ 
\hline 
\Gamma_{\rm tot}^{P^\pm(950)} [{\rm MeV}]  
& 14 & 25  \\ 
\hline 
\end{array}
\,, \label{tot:widths}
\end{eqnarray}
which shows that the $P^{\pm,3}(950)$ are quite narrow resonances.

Note 
that the $P^{\pm,3}(950)$ are basically NG 
bosons, so they do not 
couple to longitudinal modes of weak gauge bosons, 
which are essentially the NG bosons, and hence the coupling would be the forbidden three-NG-boson vertex as mentioned before, 
as far as the non-anomalous 
part with the intrinsic-parity even is concerned.  
The couplings to $WZ$ and $ZZ$, corresponding to the transverse modes,  
 then arise from the loop-induced anomalous term, the Wess-Zumino-Witten term with  
 the intrinsic-parity odd as in Eq.(\ref{P950-couplings}).  
(Note again that the $SU(8)$ symmetry forbids the coupling to $WW$.)  
Thus all the $P^{\pm,3}(950)$ couplings are necessarily loop-suppressed, hence  
the total widths are very small as in Eq.(\ref{tot:widths}). 
Thus the $P^{\pm,3}(950)$ have small couplings to weak gauge bosons,  
yielding the small $P^{\pm,3}$ production cross sections 
to easily escape from the current LHC limits, as will be seen later.

Of interest  
is that the charged $P^\pm(950)$ mainly decay to $W^\pm \gamma$ rather than $W^\pm Z$ 
(See Eq.(\ref{BR:Ppm})). 
This is simply due to the suppression by the weak mixing angle for the coupling to $Z$ compared to that to photon 
(See Eq.(\ref{P950-couplings})). 
This feature is in sharp contrast to other model isotriplet heavy Higgses  
which 
hardly decay to $W\gamma$ as addressed above. 
Hence the $P^\pm(950) \to W^\pm\gamma$ channel will give the characteristic signature at the LHC, 
a smoking gun of the one-family walking technicolor,
although the production cross section is somewhat small, 
as will be discussed below.

Now we discuss the $P^{\pm,3}(950)$ signatures at the LHC. 
First of all, we look into the neutral $P^3(950)$.  
Because of the large coupling to diphoton as in Eq.(\ref{BR:P3}),  
the $P^3(950)$ can dominantly be produced by the photon photon fusion $(\gamma\gamma {\rm F})$. 
Using the effective photon approximation~\cite{vonWeizsacker:1934nji} as in the literature~\cite{Csaki:2015vek}, 
we may calculate the production cross section of $P^3(950)$ at $\sqrt{s}=13$ TeV 
via the elastic 
photon photon fusion process to get 
\begin{eqnarray} 
\sigma^{\rm 13 TeV}_{\gamma\gamma{\rm F}} 
(pp \to P^3(950))  
\Bigg|_{\rm elastic} 
\simeq 0.018 (0.034) \,{\rm fb} 
\,, \label{ela:cross}
\end{eqnarray}  
for $N_C=3(4)$. 
Including the inelastic scattering contributions would largely enhance the cross section 
as discussed in several works listed in Refs.~\cite{Fichet:2015vvy,Molinaro:2016oix}. 
According to those literatures, the enhancement factor will be ${\cal O}(20)$, or more, 
normalized to the elastic scattering process at the resonance mass of 750 GeV.  
Quoting the result in Ref.~\cite{Molinaro:2016oix} and 
scaling the resonance mass $(m_R)$ from 750 GeV up to 950 GeV, 
one finds 
$\sigma^{13{\rm TeV}}_{\gamma\gamma F}(m_R=950\,{\rm GeV})
/\sigma^{13{\rm TeV}}_{\gamma\gamma F}(m_R=750\,{\rm GeV}) \sim 0.76$.  
Taking into account this factor together with the enhancement factor as above, 
we may roughly estimate the production cross section, 
\begin{equation} 
 \sigma^{\rm 13 TeV}_{\gamma\gamma{\rm F}} 
(pp \to P^3(950))  
\Bigg|_\textrm{elastic + inelastic} 
\sim 0.27 (0.52) \,{\rm fb} 
\,. 
\end{equation}  
Using the numbers listed in Eq.(\ref{BR:P3}) we thus estimate the 
$P^3(950)$ signal strengths: 
\begin{eqnarray} 
\begin{array}{c||c|c} 
\sigma^{13{\rm TeV}}_{\gamma\gamma {\rm F}}(P^3) \times {\rm Br} \, [{\rm fb}] & N_C=3 & N_C=4 \\ 
\hline 
\gamma\gamma 
& 0.24 & 0.46  \\ 
\hline 
ZZ 
& 0.028 & 0.052  \\ 
\hline 
Z\gamma 
& 0.00091 & 0.0015  \\ 
\hline 
\end{array}
\,. 
\end{eqnarray}
The most stringent signal is seen in the diphoton channel, which is compared 
with the ATLAS and CMS 13 TeV limits at around 950 GeV, 
$\sigma_{\gamma\gamma}^{\rm ATLAS13} \lesssim 1.6$ fb (${\cal L}=3.2$ fb$^{-1}$) 
and $\sigma_{\gamma\gamma}^{\rm CMS13} \lesssim 5$ fb (${\cal L}=2.6$ fb$^{-1}$), 
so it is far below the present bound, to be excluded, or detected in the future experiments 
with higher statistics.

We next turn to the charged $P^\pm(950)$ production at the LHC. 
 Looking at Eq.(\ref{BR:Ppm}) we find that the $P^\pm(950)$ couple  
to the diboson $WZ$, so they can be singly produced by the vector boson fusion (VBF). 
Applying the effective vector boson approximation~\cite{Kane:1984bb} with 
the parton distribution function {\tt CTEQ6L1}~\cite{Stump:2003yu}, 
we may estimate the 13 TeV production cross section of the $P^\pm(950)$ to get 
\begin{equation} 
\sigma^{\rm 13TeV}_{\rm VBF}(pp \to WZ \to P^\pm + jj) 
\simeq 0.18 (0.31) \,{\rm fb} 
\,, 
\end{equation}
for $N_C=3$(4), where $j$ denotes quarks and anti-quarks. 
 Using the numbers displayed in Eq.(\ref{BR:Ppm}) we thus calculate 
 the signal strengths of the $P^\pm(950)$: 
 \begin{eqnarray} 
\begin{array}{c||c|c} 
\sigma^{13{\rm TeV}}_{\rm VBF}(P^\pm) \times {\rm Br} \, [{\rm fb}] & N_C=3 & N_C=4 \\ 
\hline 
W \gamma + jj 
& 0.14 & 0.24  \\ 
\hline 
WZ + jj  
& 0.041 & 0.073  \\ 
\hline 
\end{array}
\,. 
\end{eqnarray}
As to the $WZ$ channel, the ATLAS Collaboration has placed the 95\% C.L. upper limit at 8 TeV (${\cal L}=20.3$fb$^{-1}$) on 
charged scalar resonances produced via the VBF, which is $\sigma_{\rm VBF}^{\rm 8TeV}(WZ) \lesssim 70$ fb at around 950 GeV~\cite{Aad:2015nfa}. 
On the other hand, the $P^\pm(950)$ predicts $\sigma_{\rm VBF}^{\rm 8TeV}(pp \to P^\pm \to WZ) \simeq 0.0088 (0.016)$ for $N_C=3(4)$, 
so it is far below the presently available upper bound.

As noted above, the $W\gamma$ cross section is much larger than the $WZ$ cross section, 
in contrast to other charged heavy scalars like in models with 
the extended Higgs sector. 
This $W\gamma$ signal is the salient phenomenological feature of the $P^\pm(950)$, 
to be tested in the future LHC experiments.

Actually, the $P^{\pm,3}(950)$ can be produced also through 
the decay of the technirho (denoted as $\rho_\Pi$), which might 
be responsible for the 8 TeV diboson excess at around 2 TeV~\cite{Fukano:2015hga}: 
the $\rho_\Pi$ couplings to the $P^{\pm,3}$ can be read off from the third reference of Ref.~\cite{Fukano:2015hga}. 
As done in the 8 TeV analysis in the references, 
we may set the overall strength of the diboson coupling $(g_{\rho\pi\pi})$ to 4 so as 
to control the total width of the $\rho_\Pi$ to be less than 100 GeV, which is fitted to the ATLAS diboson excess data~\cite{Aad:2015owa}. 
As to the Drell-Yan coupling of the $\rho_\Pi$ ($F_\rho$), however, it is now more severely constrained 
by the 13 TeV diboson data, most stringently on $WZ \to jj \nu\bar{\nu}$~\cite{ATLAS13:Jnunu}, updated from the previous publication~\cite{Fukano:2015hga}, 
to be $F_\rho \lesssim 350$ GeV. 
(The $\rho_\Pi$ diboson cross section with the Drell-Yan coupling $F_\rho \lesssim 350$ GeV cannot account for 
the 8 TeV excess, which is due to the current tension between the 8 TeV and the 13 TeV results on the diboson data.)   
Taking account of
these, we find that the branching ratio for $\rho_\Pi \to PP$ is about 3\%. 
By scaling the result in Ref.~\cite{Fukano:2015hga} we thus estimate the $P^{\pm,3}(950)$ pair production 
cross section at 13 TeV: 
\begin{eqnarray} 
&& \sigma_{\rm DY}^{\rm 13TeV}(pp \to \rho_\Pi^3 \to P^+ P^-) 
 \simeq 0.30 \,{\rm fb} 
 \,, \nonumber \\ 
 && \sigma_{\rm DY}^{\rm 13TeV}(pp \to \rho_\Pi^\pm \to P^\pm P^3) 
 \simeq 0.59 \,{\rm fb} 
\,,  
\end{eqnarray}
for $F_\rho = 350$ GeV and $g_{\rho \pi\pi}=4$. 
In this production process the final state topology will be like multiphoton plus jets through 
the dominant decay modes $P^3 \to \gamma\gamma$ and $P^\pm \to W \gamma$, 
in which two of the multiphoton are to be detected with the invariant mass around 950 GeV and 
all the final states can fully be reconstructed to be the 2 TeV resonance. 
This is an exotic topology, so would be a clean signal to be tested at the future LHC experiments.

In conclusion, 
the LHC 750 GeV diphoton excess implies the presence of yet another resonance at 950 GeV, 
that is the color-singlet isotriplet-technipion, $P^{\pm,3}$, 
in the one-family model of walking technicolor.
The $P^{\pm,3}$ mass is completely fixed at 950 GeV, which is free from any detail of the walking dynamics, 
once the 750 GeV resonance is identified with the color-singlet isosinglet-technipion, $P^0(750)$. 
The $P^{\pm,3}(950)$ are singly produced at the LHC via vector boson and photon fusion processes, 
and doubly produced by the (2 TeV) technirho decay. 
Those technipions predominantly decay to $W \gamma$ (for the charged $P^\pm(950)$) and 
$\gamma\gamma$ (for the neutral $P^3(950)$). 
In particular, the charged $P^\pm(950)$ signal is quite intrinsic for the $W\gamma$ channel, 
which yields sizable cross section, 
leading to an intriguing topology such as dijet plus mono-photon (along with forward jets). 
This is the rare signal for other charged heavy scalars as in models with the extended Higgs sector, 
so it will be characteristic only for the $P^\pm(950)$, to be accessible at the Run 2, or 3.

In addition to the color-singlet technipions, 
there are colored ones in the technipion ``zoo" in the one-family 
walking technicolor.    
As noted in the early stage of the present paper, 
colored technipion masses are predicted to be around TeV, though they are subject to 
details of the walking dynamics. 
The colored technipions would also show up in the LHC experiments, through the large signals in the dijet channel, 
or monojet and single photon, as was analyzed in the literature~\cite{Jia:2012kd,Kurachi:2014xla}. 
Thus, a number of technipions are standing by behind the 750 GeV one in the one-family walking technicolor.

More precise estimation of the walking signals in the technipion ``zoo" 
and comparison with the standard model background 
will be pursued in another publication.

In closing, in the present analysis we have so far been restricted to 
discuss the technipion couplings to the standard-model gauge bosons.  
Besides those, actually the technipions may be allowed to couple to the standard-model fermions, 
through extended technicolor interactions, 
though those couplings are formally generated at higher loops involving physics well outside of the waking technicolor dynamics. 
Among the standard model fermions, 
the Yukawa couplings to top quark and bottom quark pairs would be most influential 
to give significant corrections to 
the branching fraction of the technipions, as explicitly discussed in Ref.~\cite{Jia:2012kd}. 
The strength of such Yukawa couplings are  actually 
highly dependent on the details of the extended-technicolor model-building, 
such as the variants of strong extended technicolor~\cite{Miransky:1988gk}
having  anomalous dimension, $1<\gamma_m <2$, even larger than the walking technicolor. 
Hence we have disregarded those Yukawa couplings 
in the present analysis, in order to 
estimate effects of purely the walking technicolor dynamics as a starting point of the future analyses. 
The detailed study on the phenomenologically allowed size of the Yukawa couplings, 
and the related flavor physics predicted from the walking technicolor    
will be done elsewhere.

\acknowledgments 

We thank Ken Lane for his stimulating correspondence. 
S.M. is also grateful to Shinya Kanemura, Kenji Nishiwaki, and 
Koji Tsumura for useful comments on the technipion phenomenology 
during discussions in the 17th New Higgs Working Group, held at University of 
Toyama. 
This work was supported in part by 
the JSPS Grant-in-Aid for Young Scientists (B) \#15K17645 (S.M.).

\end{document}